\documentclass[letterpaper]{aa520}
\usepackage{txfonts}
\usepackage{graphicx}
\usepackage{aalongtable}
\usepackage{deluxetable}
\usepackage{natbib}
\usepackage{lscape}
\usepackage{epsfig}

\bibpunct{(}{)}{;}{a}{}{,} 

\newcommand{\C}[0]{ {\cal K}}

\begin{document}

\title{SALT: a Spectral Adaptive Light curve Template for Type Ia Supernovae}

\author{J. Guy, P. Astier, S. Nobili, N. Regnault, R. Pain}

\institute{Laboratoire de Physique Nucl\'eaire et des Hautes Energies,\\
           IN2P3 - CNRS - Universit\'es Paris VI et Paris VII
           4 place Jussieu
           Tour 33 - Rez de chauss\'ee
           75252 Paris Cedex 05}

\date{May 18, 2005}

\abstract{
We present a new method to 
parameterize Type Ia Supernovae (SN~Ia) multi-color light curves.  
The method was developed in order to analyze the large number of
SN~Ia multi-color light curves measured in current high-redshift projects. 
The technique is based on empirically modeling SN~Ia luminosity variations as a function of phase, 
wavelength, a shape parameter, and a color parameter. The model is trained with a sample of 
well-measured nearby SN~Ia and then tested with an independent set of supernovae by
building an optimal luminosity distance estimator that combines the supernova rest-frame luminosity,
shape parameter, and color reconstructed with the model. The distances we measure using $B$-   
and $V$-band data show a dispersion around the Hubble line comparable or lower than obtained 
with other methods. With this model, we are able to measure distances using  $U$- and $B$-band 
data with a dispersion of $0.16\pm0.05$ around the Hubble line.
}

\maketitle

\section{Introduction}
Type Ia supernovae (SNe~Ia) are a powerful tool for studying the evolution 
of the luminosity distance as a function of redshift and for subsequently constraining the 
cosmological parameters. SNe~Ia are indeed very luminous and ``standardizable'' candles, and
have lead to the discovery of the acceleration of the Universe~\citep{Riess98b,Perlmutter99}.

Although often described as a homogeneous class of objects, SNe~Ia exhibit variability in 
light curve shapes, colors, intrinsic luminosity, and spectral features. Finding 
correlations among SN~Ia observables is motivated by improving
the estimation of their intrinsic luminosity on an event-by-event basis, in
order to reduce the scatter in luminosity distance estimates. 
The main correlations observed in photometric measurements are:  
\begin{itemize} 
\item a width-luminosity (or brighter-slower) relation, which expresses the fact that 
brighter supernovae have a slower decline rate than fainter ones~\citep{
Pskovskii77,
Phillips93,
Riess95,
Hamuy96a,
Perlmutter97}.
\item a brighter-bluer relation, which was made explicit in~\citet{
Tripp98, Tripp99, Parodi00}, and assumed to be due to 
extinction by dust in other works \citep{
Riess96a,
Riess98b,
Perlmutter99,
Tonry03,
Knop03,
Barris04}. 
\end{itemize}
The case of the brighter-bluer relation is interesting.
Even if authors fully agree neither on the origin of the effect nor on the 
strength of the correlation, most, if not all, recent attempts to build a SN~Ia Hubble
diagram have made use of color in their distance
estimator\footnote{In \citet{Perlmutter99}, it is checked that $B-V$
rest-frame colors of nearby and distant type agree on average, and the
color measurement is not used event per event.}.
\citet{Riess96b} summarizes previous work on the subject and proposes
a way to reconcile divergent interpretations of data by taking into account
the correlation between light curve shape and color.
New methods using color information have also been recently 
proposed to estimate luminosity
distances (see for example \citealt{Wang03} and \citealt{Wang05}).
We will bring to the
debate our own estimations of the correlation strength and its
wavelength dependence.

Cosmological measurements using SNe~Ia are based on comparing
nearby and distant objects. In order to reduce the sources of
systematic uncertainties, it is important that all distances are
derived using the same procedure, especially when considering 
large samples such as those being collected by the 
ESSENCE\footnote{{\tt http://www.ctio.noao.edu/\~{ }wsne}} or 
SNLS\footnote{{\tt http://cfht.hawaii.edu/SNLS}} projects. To analyze
these data sets, the following constraints have to be taken into
account:
\begin{itemize}
\item High redshift objects often lack late-time photometric data 
(or have one of too poor quality) 
which makes it impracticable to estimate color from late-time 
data as proposed by
\citet{Lira98}.
\item 
Above redshifts $z\sim0.8$, 
rest-frame $U$-band measurements have to be used 
because of the limitation of silicon
detectors.  
Incorporating rest-frame $U$-band measurements on the same footing
as the commonly used rest-frame $B$- and $V$-band data is hence highly
desirable.
\item Applying cuts on light curve parameters should be prohibited since 
the resolution on these parameters follows the photometric resolutions and
hence degrades with redshift, therefore biasing the event sample.
In particular, cuts on
colors to eliminate reddened events are to be avoided.
\end{itemize}

Various techniques have been used to estimates distances.
\cite{Phillips93} and
\cite{Perlmutter97} fit one band at a time, and derive distances
from light curve parameters measured in different bands (usually $B$ and
$V$ bands), sometimes relying on late time measurements to measure host
galaxy extinction. A recent and refined version of this approach can
be found in \cite{Wang05}, and in \cite{Wang03} it is proposed to fit $B-V$
as a function of $B$. These methods do not make use of rest-frame $U$-band
measurements. 
\cite{Nugent02}
discuss in detail the problems associated with SN~Ia
$U$-band photometry. They mention in particular the apparent large intrinsic
variations of the UV luminosity among similar supernovae as well as
our lack of understanding of the SN~Ia UV photometry --principally due
the uncertainties which affect the large extinction corrections which
have to be applied to the data. Nevertheless, it is important to try and
incorporate rest-frame $U$-band data to estimate distances.

An extension of \cite{Perlmutter97} to rest-frame $U$-band data was
developed in \cite{Knop03}, but at the expense of adding a large distance
systematic (and probably statistical) uncertainty. Similarly, the MLCS method
\citep{Riess95,Riess96a} has been extended to 
include $U$-band measurements
under the name of MLCS2k2 \citep{Jha02} and used in \cite{Riess04}, but 
also at the expense of a worsened distance resolution \citep{Jha02}. 
We will show that our model predicts
light curves for any band located between rest-frame $U$ and $R$ bands and 
that we are
able to get a similar or better distance resolution  
with both rest-frame ($B,V$) and ($U,B$) band pairs. 

We propose here to parameterize the
light curve model with a minimal parameter set: a luminosity parameter,
a decline rate parameter and a single color parameter. Our approach
will be to build a phenomenological model of the expected SN flux,
continuously varying with phase, wavelength, decline rate and color,
in order to capture all these features at once. 
This approach offers
several practical advantages which make it easily applicable to
high-redshift SNe~Ia currently measured in large projects. 
First, the k-corrections are built into the model and not applied to
the data. This allows one to propagate all the uncertainties directly
from the measurement errors.
More importantly, when needed, we make use of the SN rest-frame 
$U$-band fluxes to estimate the supernova distances. 

In Sect. 2 we describe the semi-analytic model used. 
We then describe, in Sect. 3, how the coefficients of the model are determined 
by an iterative training based
on a set of well-sampled nearby SNe~Ia taken from the literature. We also
highlight some properties of the resulting model. At this stage, the aim 
is to model  multi-color light curves and not to estimate luminosity distances.
The model is tested in Sect. 4 with an independent set of SNe~Ia in the Hubble flow.
A luminosity distance estimate is then constructed from the fitted parameters of the 
light curve model. It is used 
to build Hubble diagrams successively from $(B,V)$ and $(U,B)$ 
light curve pairs.
In order to assess the precision of this approach, we compare
distance estimates of the same events obtained from 
the $(U,B)$ and $(B,V)$ band pairs and 
with other distance estimators.

\section{The light curve model}
\label{section:model}

\subsection{Model definitions}
As already mentioned, we choose to parameterize light curves (more precisely
light curve pairs or light curve triplets when available) using a single 
luminosity, 
a single shape parameter and a single color.
The choice among possible implementations is largely arbitrary. We 
choose the following parameters, which enable comparisons to be made with previous works:
\begin{itemize}
\item ${\bf f_0}$ : a global intensity parameter which varies with redshift like 
the inverse of the luminosity distance squared,
\item ${\bf s}$ : a time stretch factor as the decline rate indicator \citep{Perlmutter97}. 
In \citet{Goldhaber01}, this parameter is shown to apply to the 
rising part of the light curve as well. However, while the stretch paradigm 
describes
well the bright part of the $B$ light curve, it does poorly
at late time. It also fails to capture 
the shape variations in the other bands. This is why our model
uses the stretch parameter as an index rather than the stretch paradigm itself.
As described below, by construction our model follows exactly the 
stretch paradigm in the $B$ band.
\item {${\bf c}$} = $(B-V)_{max} + 0.057$, where $(B-V)_{max}$ is measured 
at B maximum,  and $-0.057$ is the chosen reference color (Vega magnitudes) 
of a SN~Ia.
\item $t_{max}^B$ :  the date of maximum in the rest-frame $B$ band  
\end{itemize}
With these definitions, the expected counting rate $f_{SN}$ in a given pass-band $T$, of a supernova 
at redshift $z$, and at a phase $p\equiv(t-t_{max}^B)/(1+z)$, can be written:

\begin{eqnarray}
f_{SN}(p, z, T) = f_0 \, (1+z)  \int \phi(p, \lambda, s, c) \ \frac{\lambda}{h\, c}\ T(\lambda(1+z)) \,  d\lambda \nonumber
\end{eqnarray}
$\phi(p,\lambda,s,c)$ is a model of the SN~Ia energy luminosity per
unit wavelength. It may vary with the supernova stretch and color.
Note that the potential extinction by dust in the host galaxy is not explicit
in the equation. Instead, we choose to incorporate it in the model $\phi(p,\lambda,s,c)$
as discussed below. 

Building an
average spectral template $\phi$ as a function of phase, wavelength, color and
stretch from observations is complicated because of the inhomogeneity and 
incompleteness of published
data. Although some spectral features have been correlated with
stretch, we do not have yet a complete knowledge of spectral
diversity as a function of phase. In most of the approaches to SN~Ia
light curve fitting, the limited knowledge of spectral variability
impacts on the accuracy of cross-filter k-corrections, defined as
expected ratios of fluxes in different bands at the same phase
\citep{Kim96, Nugent02}. 

In \cite{Nugent02}, it was shown that the variation in 
k-corrections from one supernova to another depends 
primarily on the supernova color, 
and to a lesser extent on spectral
features\footnote{Most of the SN~Ia photometric reductions transform
instrumental magnitudes into standard magnitudes using color
equations, derived from standard stars observations. This assumes
that color rather than spectral features dominates cross-filter
corrections.}. 
Hence we neglected the variability of those spectral features   
in the modeling of light curves.

In order to implement stretch dependent light curve shapes and colors
we therefore used the following approximation:
\begin{eqnarray}
f_{SN}(p_s,z,T) = f_0 \, (1+z)  \int \phi(p_s, \lambda) \, \frac{\lambda}{h\, c} \, T(\lambda(1+z)) \,  d\lambda \nonumber\\ 
\times  \  \exp \left[ -0.4 \, \ln(10) \times \C (p_s, \lambda_T, s, c) \right]
\label{eq:model}
\end{eqnarray}
where $p_s \equiv p/s$ is a stretch-corrected phase. This functional
 form defines the light curve model.

In equation~\ref{eq:model}, $\phi$ no longer depends explicitly on $s$ and $c$, and
 $\C(p_s,\lambda,s,c)$ is a smooth ``correction'' function of our four 
variables. $\lambda_T$ is the central wavelength of the filter
 $T$. $\C$ enables one to implement light curve shape variations 
that are more complicated than simple dilation of the time scale, 
 along with stretch dependent colors. 
As described below, $\C$ varies smoothly 
  with $\lambda$; this justifies placing it outside the 
integral over wavelength. We also 
considered 
keeping $\C(p_s, \lambda_T, s, c)$ inside 
the integral. It changed the results of the fits by less than 1\%, 
while the computing time was multiplied by a factor of 10.

Equation (\ref{eq:model}) implements
k-corrections for an average SN~Ia, conforming to the common
practice. The approach proposed here will be easy to
adapt when constructing a stretch-dependent spectral
template becomes possible.

  The k-corrections are usually applied to data. Here, they are 
incorporated into the model. This offers a few practical advantages:
first, light curves can be generated for arbitrary pass-bands (within
the spectral coverage of $\phi$ and $\C$); light curves in the observed
pass-bands can be directly fitted to the data.
Second, the light curve parameter
uncertainties extracted from the fit incorporate all uncertainties
propagated from measurement uncertainties; for example, uncertainties
introduced in the k-corrections by the possibly poor determination
of the date of maximum and/or color are propagated into the
parameters uncertainties. 

The functions $\phi$ and $\C$ define the model. Once they are determined, one can fit 
the supernova photometric data points, measured in a minimum of 
two pass-bands, 
to estimate $f_0$, $s$, $c$, and a date of $B$ maximum light, 
which is a nuisance parameter. With only one passband, $c$ must be held fixed.

For $\phi(p_s,\lambda)$, we use a template spectrum 
assembled by P. Nugent~(\citealt{Nugent02} and private communication)
smoothed along the phase (time) axis, and normalized as a function of phase 
to the $B$-band light curve template ``Parab -18'' of \citet{Goldhaber01}. 
Any smooth variation of the template $\phi$ with phase or wavelength is irrelevant 
at this level, since it may be changed by the function $\C(p_s,\lambda,s, c)$. 
The only important quantities in $\phi(p_s,\lambda)$ are the SN~Ia spectral features, 
intended to be realistic on average.

The empirical correction function $\C$ is implemented as a sum of two polynomials:
\begin{equation}
\C(p_s,\lambda,s,c) = \C_s(p_s,\lambda,s) + \C_c(\lambda,c) \label{eqn:correction_functions}
\end{equation}
where we explicitly separate the corrections associated with the parameters $s$ and $c$ to 
clarify their interpretation.
$\C_s(p_s,\lambda,s)$ modifies the shape of light curves and absorbs 
any stretch--color relation except for the $(B-V)_{max}$ color.
Indeed we want $c$ to describe exactly the $(B-V)_{max}$ color. 
$\C_c(\lambda,c)$ is then a color correction as a function of wavelength and color.

In order to remove all degeneracies among coefficients, the model 
must fulfill the following constraints:
\begin{eqnarray}
\C_s(p_s<35,s,\lambda_B)          &=& 0 \nonumber \label{constraint:B0}\\
\C_s(p_s=0,s,\lambda_V)          &=& 0 \nonumber \label{constraint:Vs0}\\
\C_c(\lambda,c=0) =  \C_c(\lambda_B,c)             &=& 0 \nonumber \label{constraint:C0}\\
\C_c(\lambda_B,c)- \C_c(\lambda_V,c) &=& c  \nonumber \label{constraint:BV0}
\end{eqnarray}
where $\lambda_B$ and $\lambda_V$ refer to the mean wavelengths of the B and V filters.
The first constraint ensures that the parameter $s$ actually defines the stretch 
in $B$ band, since the template is not modified for $\lambda = \lambda_B$. The 
remaining constraints define $c$ as the $B-V$ color (relative to the spectral 
template) at maximum $B$, for all stretches. Note that other colors may 
(and in fact will) depend on $s$ at fixed $c$. 
$f_0$ describes the actual peak flux in $B$. It is in no way ``corrected''
for the brighter-slower relation.
 
$\C_s$ is implemented as a polynomial of degree $D_p$ in phase,
$D_\lambda$ in $\lambda$ and $D_s$ in stretch respectively. Similarly,
$\C_c$ is a polynomial of degree $D_p$ in phase and $D_c$ in color. In
this study, we chose for the degrees of polynomials $(D_p, D_\lambda,
D_s, D_c ) = (4 ,3 ,1, 1)$. The large degree in phase permits a
detailed adaptation of the model light curve shapes to the actual
data. The other degrees correspond to the minimal number of
coefficients (a degree 3 in wavelength is chosen to adapt colors in
$UBVR$, independently). The $\C$ polynomials are then defined by 48 
coefficients, but
the constraints reduce the number of independent coefficients to
34. There are 32 free coefficients for $\C_s$ and 2 free 
coefficients for $\C_c$.  We
will call ``training'' the determination of these coefficients from
measurements of nearby SNe~Ia.

\subsection{Normalization of transmissions}

The value of  $f_0$  depends on the normalization of the spectral 
template $\phi$. This is not an issue
when comparing nearby and distant supernovae to measure cosmological
parameters, since only flux ratios matter: all objects have to be
fitted using the same model. For $H_0$
measurements, one would need to normalize $(\phi)$ using SNe~Ia at known
distances, but we will not perform this here.

More important is the apparent dependence of the value of $f_0$ with respect 
to the normalization
of the transmission $T$ (equation~\ref{eq:model}). Since we aim at fitting multi-color 
light curves, observed with different 
instruments, with a single set of parameters $(f_0,s,c)$, it is mandatory to 
remove this dependence.

All photometric data are expressed in units of the integrated flux $f_{ref}$ 
 (deduced from the zero point) of a known standard spectrum $\phi_{ref}$ (e.g. Vega).
While the functional form of $T(\lambda)$ is 
     determined by optical transmission measurements its normalization
     can be determined from $f_{ref}$ and $\phi_{ref}$ via the relation
\begin{equation}
\int \phi_{ref}(\lambda) \, \frac{\lambda}{ h\, c} \,  T(\lambda) \, d\lambda \,=\, f_{ref} \label{eq:fref} 
\end{equation}

The filter transmissions are usually published as series of numbers 
\citep[as in][]{Bessel90}, and this leads to ambiguities: for example,
\cite{Suntzeff99} writes {\it "Note that \cite{Bessel90}
defines the sensitivity function as the product of the quantum
efficiency of the detector+telescope, the filter transmission curve,
the atmospheric extinction, and a linearly increasing function of
wavelength."}. The ambiguity is whether the transmission refers to
signal per unit energy, or signal per photon flux. 
Attributing dimensions to transmissions solves the ambiguity.

There have also been concerns about the relative weights of the 
different wavelengths in the integrals over wavelength \citep{Nugent02},
namely whether one should sum photons or energies.
If one aims at reproducing the instrument response to an arbitrary 
spectral energy density, the mathematical integrals should mimic 
the physical integration process of the instrument \citep{Fukugita96}.
For example, if one considers a CCD-based
observing system, the effective transmission $T(\lambda)$ 
will be proportional to 
a number of photo-electrons per photon. 
So one should not integrate photon counts
nor energies, but charge on the detector (or ADC counts).

\section{Training the model}
Since $f_0$ describes the {\it observed} luminosity in the
$B$ band (because of the constraints applied to the corrections), the
model only incorporates stretch-shape and stretch-color relations, but
no correlation involving luminosity. This option was chosen in order
to allow us to train the model with objects at unknown distances, in
particular the nearby objects in the sample of \citet{Jha02} measured
in the $U$ band.  If one offsets all magnitudes of each training
object by an arbitrary amount, possibly different for each object, the
resulting model will not change.  One could then consider incorporating
high redshift objects into the training, 
but we choose not to do it here.

\subsection{The nearby SN~Ia Sample}
The model was trained and tested using a sample of published nearby
supernova light curves. We collected 122
SNe~Ia for which $B$- and $V$-band light curves are available in
the literature, including data from \citet{Hamuy96a,Riess96b} and
\citet{Jha02} for a total of 94 objects, and 28 additional supernovae
collected from various sources (see the caption of
table~\ref{tab:training_sample}).

Objects were then selected based on two main criteria.
First, we kept 
supernovae with at least two measurements before
the maximum in the $B$ or the $V$ band. This is necessary to ensure
that the date of maximum is well defined and that the measurements
can safely be used as a function of phase. 
Out of the whole sample, 56 SNe satisfied this criterion. Then 
under-luminous peculiar supernovae:
SN~1991bg,
SN~1998bp, SN~1998de and SN~1999by \citep{Howell01,Li01b}, and the
peculiar objects SN~2000cx and SN~2002er
\citep{Li01c,Pignata04} were rejected from our sample.
There are a number of reasons for this
choice; the most important one being the spectral difference between normal
and SN~1991bg-like SNe~Ia. Since our model is built using a spectral
template describing the features of the average normal SN~Ia, it is
not well suited for describing very different objects, such as
SN~1991bg-like events. We note that, this is not a constraint for
cosmological studies since, so far, no under-luminous SNe have been
found in distant SN searches. Furthermore, if present at high-redshift, 
these objects would easily be 
identified both from spectroscopy and from their $B-V$ color 
evolution \citep[see for instance][]{Phillips99,Garnavich04}.

SN~1991T-like events were kept in 
the sample. This "sub-class" of over-luminous SNe~Ia is
spectroscopically identified by the presence of unusually weak
absorption lines during the pre-maximum phase. These are more
difficult to identify than their under-luminous counterparts in low
signal-to-noise spectra, such as those usually available for high
redshift SNe.
Moreover, the $B-V$ color evolution is not particularly
different from the one of normal SNe. Ideally, one should use a
different spectral template for fitting this kind of events. We
estimate the errors introduced in the k-corrections using a standard
 SN~Ia spectral template 
to be of a few percent in the
worst cases, and well within the final dispersion in the peak
luminosity of SNe~Ia. We further note that, the uncertainty due to 
using a non optimal spectral template is analogous to the
uncertainty in the cross filter $k$-corrections used in standard methods.

The resulting sample of 50 SNe was then split into two sets: a training
sample and a test sample.  The training sample was used to adjust the
coefficients of the polynomials of the model. It contains all the
supernovae with redshifts smaller than 0.015 (not in the Hubble flow) 
and 6 supernovae at
redshifts above 0.015, for which with $U$-band data was available, in order 
to improve the model in this wavelength region. The training sample 
contains the 34
supernovae listed in table~\ref{tab:training_sample}.
The test sample contains 26 supernovae (table~\ref{tab:test_sample}).
Note that the two samples are not completely independent. Indeed, they
share ten supernovae with $U$-band light curves, since such events are
rather scarce.

The data was not pre processed in any way prior to fitting. To account
for the Milky Way extinction, we incorporate it into the instrument
transmission, using the law from \citet{Cardelli89} with a color
excess $E(B-V)$ obtained from \citet{Schlegel98} dust maps at the
position of the object to fit.

\subsection{Training the Model}

 All the published nearby supernova magnitudes are expressed in the
 Johnson-Cousins $UBVR$ system. In Equation (\ref{eq:model}), we use models
 of the instrument transmissions as a function of wavelength. We
 adopted the transmission functions published by \citet{Bessel90},
 and interpreted them as $\lambda \, T(\lambda)$ (see equation~\ref{eq:model}), 
 i.e. counts per unit energy, following a footnote of \citet{Suntzeff99}.

 Training the model consists in determining the $\C(p_s, \lambda, s, c)
 = \C_s(p_s, \lambda, s) + \C_c(\lambda, c)$ correction function
 (Eq. \ref{eqn:correction_functions}) using the training sample data
 in the $UBVR$ bands. The $R$ band was introduced to avoid relying 
on extrapolation for measurements beyond rest-frame $V$ band. 
We start with a first guess:
\begin{eqnarray}
  \C_s(p_s, \lambda, s)    &=& 0 \nonumber\\ 
  \C_c(\lambda,c)          &=& c \times (\lambda-\lambda_B)/(\lambda_V-\lambda_B) \nonumber
\end{eqnarray}
 and we use an iterative algorithm which can be sketched as follows:
\begin{enumerate}
 \item Fit the light curves using the current determination of $\C$.
 \item Fit $\delta \, \C(p_s, \lambda, s, c)$, an instance of the $\C$ function, on the light curve residuals.
 During this step, identify and remove the outliers data points.
 \item $\C \leftarrow \C + \delta \, \C$.
 \item {\tt GOTO} step 1, until $\delta \, \C$ becomes negligible.
\end{enumerate}

 All the available data, {\em i.e.} the $UBVR$ residuals, up to large
 phases were used to determine the $\C$ function ({step
 2}). However, not all the data points were used to fit the light
 curves ({step 1}):
the $R$-band data was not used 
and
only $UBV$ photometric points with phases ranging from $-15$ days to
$+35$ days were used since we are mostly interested in describing the
rest-frame $UBV$ central part of the supernova light curves.

\subsection{Results of the training}\label{section:results_training} 

\begin{figure}
\centering
\epsfig{figure=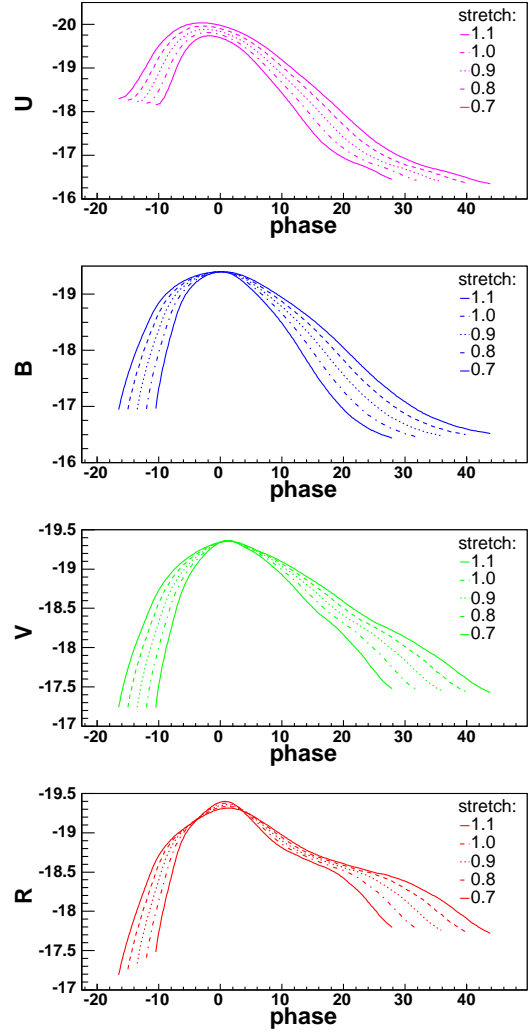,width=0.4\textwidth}
\caption{The $UBVR$ template light curves obtained after the training
phase for different values of the stretch and null color. 
Note the strong variations of the $U$-band maximum with
the stretch. 
Note also the after max shoulders in the $V$ and $R$ bands.
The model also reproduces well the fact that 
the brighter-slower relation is weaker in the $R$ 
band than in the $B$ or $V$ bands. 
The model does {\it not} incorporate the brighter-slower correlation.  
The apparent brighter-slower correlation for the $U$ band is, in fact, a stretch-color correlation.
\label{fig:templates}} 
\end{figure}

\begin{figure}
\centering
\epsfig{figure=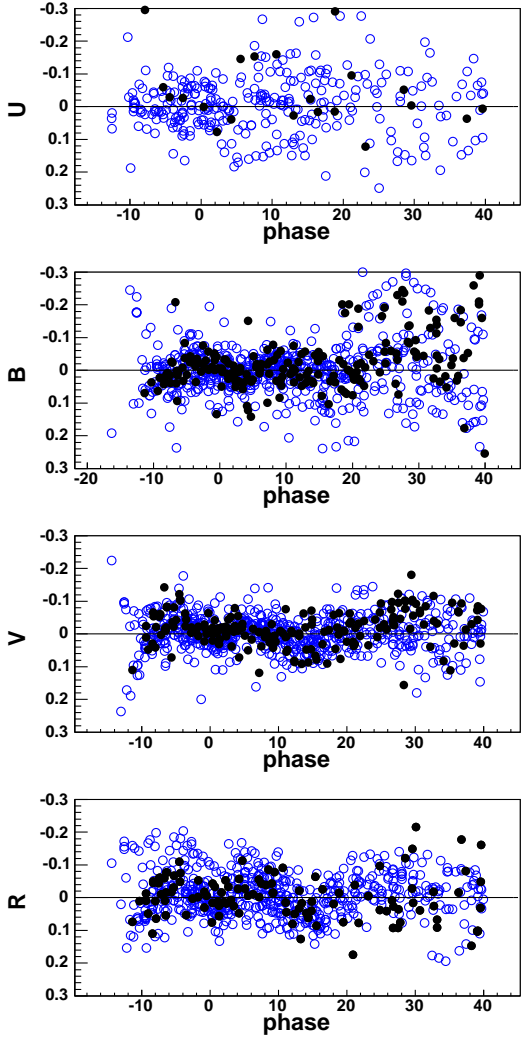,width=0.4\textwidth}
\caption{Residuals to the light curve fit as a function of phase, for supernovae from the training sample (open symbols) and the test sample (filled symbols).
There is no significant difference between the distributions obtained with both samples which proves that the model is not over-trained.
The systematic residuals as a function of phase are a direct consequence of the limited number of parameters (5) used to implement the corrections $\C_s$ as a function of phase. Increasing this number would remove this systematic trend in residuals, but some parameters would be poorly constrained in the $U$ band where photometric data is scarce.
\label{fig:residuals_lc}} 
\end{figure}

The fit converged after four iterations.
2480 measurement points were fitted, and 39 were discarded as 
outliers (at the 3 $\sigma$ level). Compared to the number of free coefficients 
of the model, we can safely conclude that the model is not over-trained.
The standard deviations of the residuals
to the model in $UBVR$ are respectively of $0.09, 0.09, 0.06, 0.07$ magnitudes. 
Figure (\ref{fig:templates}) shows the final $U$, $B$, $V$ and $R$ templates
obtained at the end of the process as a function of stretch. 
By construction, the rest-frame $B$ and $V$-band magnitudes at maximum 
do not vary with stretch. 
We find a strong dependence of $(U-B)_{max}$ with 
stretch ($\delta(U-B)_{max} \simeq -\, \delta s$, compatible with \citealt{Jha02}), 
which is an essential feature for the model to reproduce in order to 
estimate a reliable color in the wavelength range 
between $U$ and $B$. The model also manages to reproduce a 
a stretch-dependent secondary shoulder in the $R$ band. We also notice
that the $R$-band light curves cross each-other for different values of the
stretch. This reproduces well the fact that the brighter slower relation
is weaker in the redder pass-bands, as noted by \cite{Phillips99}.
The residuals to the light curve fit are shown figure~\ref{fig:residuals_lc}.

\begin{figure}
\centering
\epsfig{figure=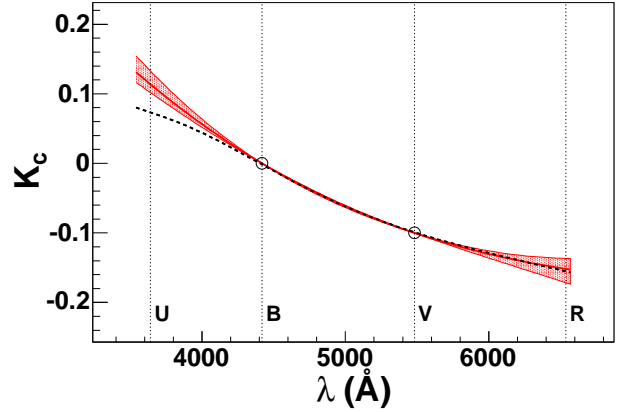,width=0.5\textwidth}
\caption{The color correction $\C_c$ as a function of wavelength for a value 
of $c$ of 0.1 (thick line). The shaded area corresponds to one standard deviation 
to the best fit value. The dashed curve represents the extinction with respect 
to $B$ band, $(A_{\lambda}-A_B)$, from~\citet{Cardelli89} with $R_V=3.1$ 
and $E(B-V)=0.1$. The curves are compatible in the $R$ band but not in 
the $U$ band (by construction they match in $B$ and $V$ bands as illustrated with
the open circles).
 \label{fig:colorcorrection}} 
\end{figure}

Figure~\ref{fig:colorcorrection} represents the color correction 
$\C_c(\lambda,c)$ for $c=0.1$ compared to the dust extinction law 
from~\citet{Cardelli89}. Interestingly enough, the law we obtain follows 
pretty well that of Cardelli in the $R$ band but not 
in the $U$ band where we get a stronger dependence on $c$. Also 
shown (shaded area) is the uncertainty on $\C_c$ derived from the $\chi^2$ 
increment (normalized to the number of supernovae) 
of the fit to the light curve residuals.
The Cardelli law in the $U$ band is at 3.7 standard deviation from the best fit value.

As a consequence, we deduce that the relation between E(B-V) and E(V-R) are
very similar (i.e. indistinguishable) to the ones expected from
reddening by dust. This similarity, noted by \cite{Riess96b}, does not prove however 
that $c$ can be interpreted as reddening by dust; an additional requirement for this 
hypothesis to be valid would be that the peak $B$-band magnitude increases with $c$ 
by a value of $R_B \times c$. We will see that this is not the case in the next 
section. 
Let us also emphasize  that the stretch dependent part of the $U-B$ and $V-R$ 
colors are included in the stretch dependent term $\C_s(p_s,\lambda,s)$, and
{\it not} in the color curve of figure \ref{fig:colorcorrection}.

\section{Performance study}
\label{section:performance_study}
Once the correction function $\C(p_s,\lambda,s,c)$ is determined, we can
fit the model on the sample of nearby SN~Ia light curves 
listed in table~\ref{tab:test_sample}.
This allows us to perform various consistency checks, in order to make
sure that the model describes well the $UBV$ photometry of SNe~Ia. 

A first test consists in checking the ability of the model to reproduce the 
shape and color features of the independent set of SNe~Ia. This is
demonstrated on figure~\ref{fig:residuals_lc} which shows the residuals to 
the fits of light curves of SNe from the test sample.

\subsection{Distance estimate}

The global intensity parameter $f_0$ is proportional to $d_L(z)^{-2}$ 
and inversely 
proportional to the normalization of $\phi$. One can define a rest-frame $B$ 
magnitude $m^*_B$~\citep{Perlmutter97} which removes this artificial 
dependence on the model normalization, 
\begin{equation}
m^*_B = -2.5 \, \log_{10} \frac{f_{SN}(0,z,T^*_B)}{(1+z) \, f_{ref}(T_B)}
\end{equation}
where $f_{SN}$ and  $f_{ref}$ are respectively  defined by equations~\ref{eq:model} 
and~\ref{eq:fref}, $T_B$ is the transmission of the $B$ filter and \\
$T^*_B(\lambda)=T_B(\lambda/(1+z))$ is a redshifted $B$ transmission. One can check 
that $m^*_B$ varies as $5 \log_{10} d_L(z)$ with redshift and that  $m^*_B \rightarrow
m_B$ for $z \ll 1$, where $m_B$ is the conventional $B$ magnitude.

We incorporate the Hubble parameter dependence of $d_L$ in a constant parameter 
$M_B^{70} = M_B - 5 \, \log_{10}\left( h_{70}\right)$, which is the average absolute 
magnitude of a SN~Ia with $s=1$ and $c=0$, for a value of the Hubble 
parameter of 70 km.s$^{-1}$.Mpc$^{-1}$.

As mentioned in the introduction, the peak luminosity of SNe~Ia is 
correlated to stretch and color, so we may build a distance estimator 
that accounts
for those correlations and as a result reduces the dispersion.
Following~\citet{Tripp98}, we adopt linear corrections of 
coefficients $\alpha$ and $\beta$ respectively 
for stretch and color. The distance estimator is then
\begin{equation}
 m^*_B - M_B^{70} - 43.16 + \alpha \, (s-1) - \beta \, c
\label{eq:hubble}
\end{equation}
Its expectation value for a supernova at redshift $z$ is $5\, \log_{10} \left( d_L(z) \, H_0 c^{-1} \right)$. 
Since our goal is here to test the distance estimator rather than actually perform 
a cosmological fit, we impose the ``concordance'' cosmological parameters 
($\Omega_M=0.3$ and $\Omega_\Lambda=0.7$) when fitting $M_B^{70}$, $\alpha$ and $\beta$ and
apply the method to build low-z Hubble diagrams using 
successively $(B,V)$ only and $(U,B)$ only light curves of supernovae from the test sample 
(table~\ref{tab:test_sample}). In general, rest-frame $(U,B,V)$ light curve triplets  
should be used, when available, to determine the light curve parameters.

\subsubsection{Hubble diagram in $BV$ only} \label{sec:hubblediagram_BV}

\begin{figure}
\centering
\epsfig{figure=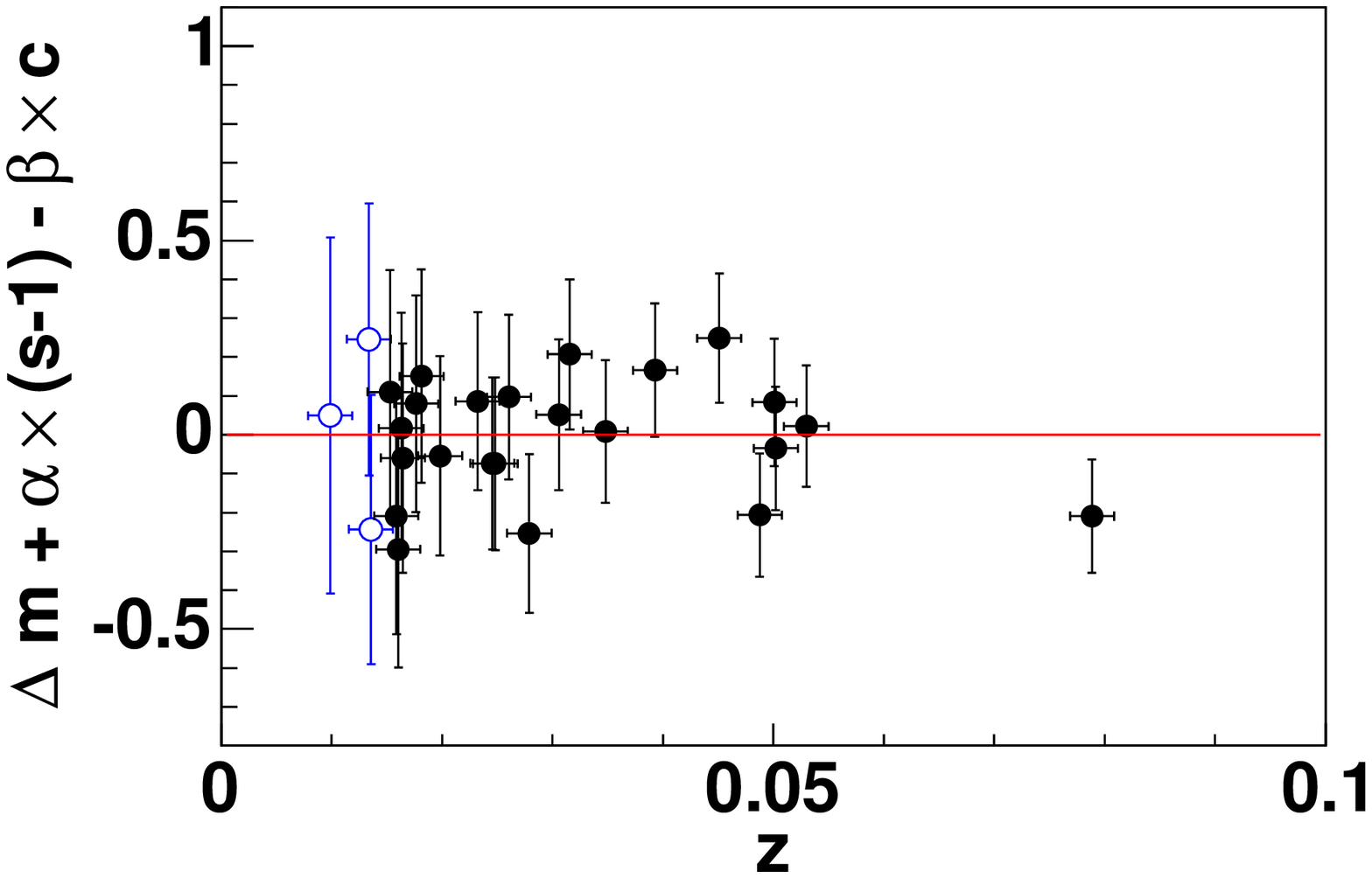,width=0.4\textwidth}
\epsfig{figure=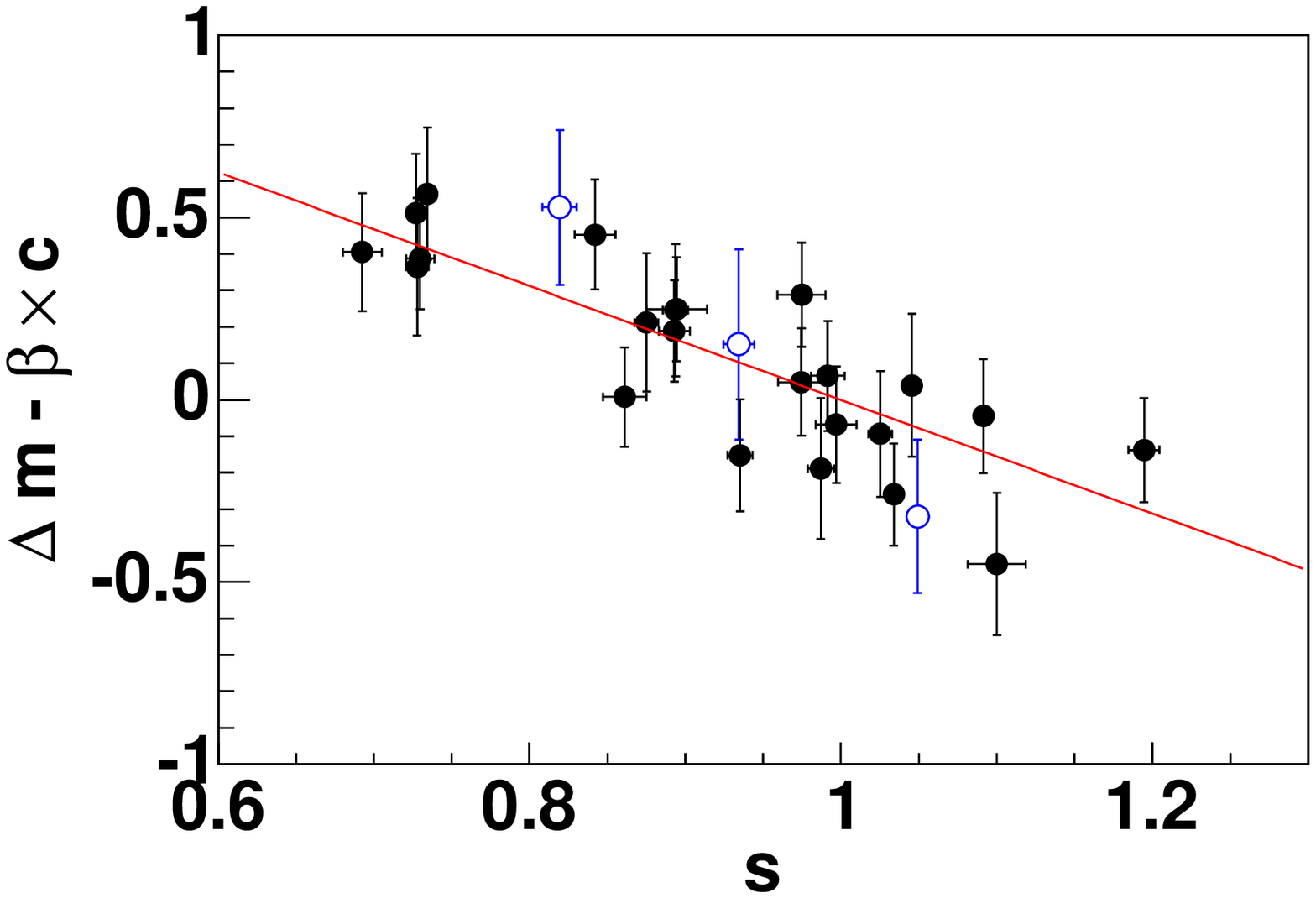,width=0.4\textwidth}
\epsfig{figure=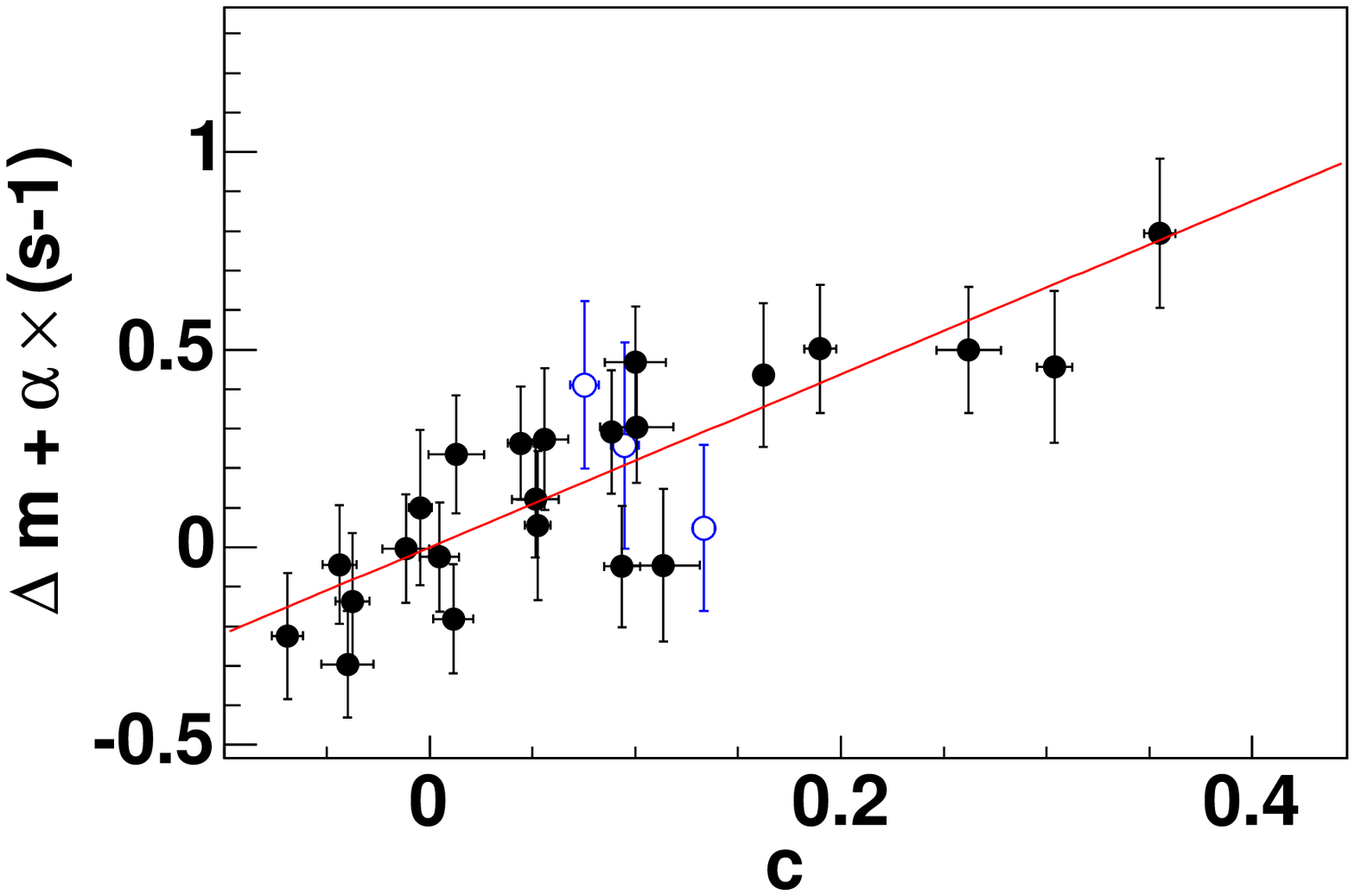,width=0.4\textwidth}
\caption{From top to bottom: Residuals to the Hubble diagram as a function of 
redshift, stretch $s$ and color $c$ indexes for supernovae of the test sample 
fitted in $B$ and $V$ bands. SNe with redshifts smaller than 0.015 are labeled 
with opened symbols.\label{fig:hubbleBV}}
\end{figure}

Using $B$- and $V$-band only light curves of supernovae 
with redshifts larger than 0.015 from the test sample, we obtain:
 $M_B^{70}= -19.41 \pm 0.04$, $\alpha= 1.56 \pm 0.25$ and $\beta= 2.19 \pm 0.33$. 
The standard deviation of residuals is $0.16 \pm 0.03$~\footnote{Note that this number takes into 
account the number of parameters in the fit. The measured {\it RMS} value is $0.14 \pm 0.03$.}. 
Uncertainties on $m_B,s,c$ along with their covariance were included in 
the fit\footnote{The uncertainties on the distance estimate formally depend on 
$\alpha$ 
and $\beta$, and increase with them. As a consequence, the $\chi^2$ minimum is 
biased toward large values of these parameters. We therefore 
computed the uncertainties with the initial values, and use the result of the fit 
 at the final iteration.}, we also considered an uncertainty on 
redshifts due to peculiar velocities of 300~km.s$^{-1}$; an additional 
``intrinsic'' dispersion of 0.13 
is needed in order to get a $\chi^2$ per degree of
freedom of 1. 
The resulting Hubble diagram as well as the 
observed brighter--slower and brighter--bluer relations are shown 
figure~\ref{fig:hubbleBV}.

   Our approach to estimating distances easily compares to the 
one adopted in \citet{Tripp98}: the main differences are the light curve
model and the brighter-slower parameterization. When we fit
the same SNe sample (The Cal\'an-Tololo sample from \citealt{Hamuy96b}),
a value of $\alpha = 1.04 \pm 0.24$ and $\beta=2.08  \pm  0.27$ are obtained, which
compare well to $\alpha = 0.88$\footnote{$b=0.52$ translates to 
$\alpha\simeq 0.88$ when using stretch and the first order 
relation $(\Delta M_{15}-1.1) \simeq 1.7 (1-s)$.}, $\beta = 2.09$ of \cite{Tripp98},
based on peak luminosity, color, and decline rate estimates from 
\cite{Hamuy96b}. We conclude that our model correctly reproduces
basic parameter estimations of previous works. 

Concerning the interpretation of the brighter-bluer correlation, 
we find a value of $\beta$ which is
incompatible with $R_B=4.1$, expected for extinction by dust analogous
to the observed law in the Milky Way. The value we find is compatible with 
those found in previous works (see \citet{Tripp98} and references therein). 
However, as stressed in \citet{Riess96b}, 
the color excess (or deficit) at maximum should not be interpreted 
as entirely due to extinction but be corrected for the 
part of this excess that is correlated with stretch . We measure
a stretch-color slope of about 0.2, similar to the relation
proposed in \citet{Phillips99}\footnote{The proposed 
relation is $\frac{dc}{d\Delta M_{15}} = 0.114 \pm 0.037$. With the approximate 
relation $\frac{d\Delta M_{15}}{d s_B} \simeq -1.7$ (at $s_B$ = 1), 
we expect $dc/ds \simeq -0.2$.}
and can redefine our parameters to account for this correlation:
\begin{eqnarray}
c' &=& c + 0.2 \,(s-1) \nonumber \\
s' &=& s \nonumber
\end{eqnarray}
so that $s'$ and $c'$ are uncorrelated. The correlation coefficients
then become $\alpha' = \alpha + 0.2 \, \beta$ and $\beta' = \beta$, which
means that   
redefining the color excess to explicitly assign to stretch the 
color variations correlated to stretch does not change 
the brighter-bluer correlation strength.

The two-parameter (stretch and color) pragmatic approach we followed
can accommodate both reddening by dust and any intrinsic color effect
dependent or not on stretch. One may reasonably assume that
reddening by dust and stretch independent intrinsic colors mix 
(as proposed in \citealt{Nobili03}), and
that disentangling the contributions would improve the distance
resolution. Our distance indicator is however independent of the
interpretation of the color variations. Since the low value of
$\beta$ may indicate that some intrinsic effect plays a role, we did
not interpret color as only due to reddening by dust and hence
accepted negative $c$ values as such.

\subsubsection{Hubble diagram in $UB$ only} \label{sec:hubblediagram_UB}

\begin{figure}
\centering
\epsfig{figure=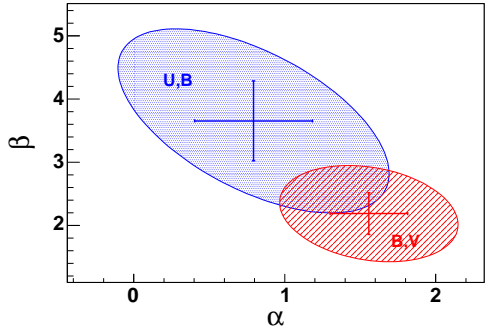,width=0.4\textwidth}
\caption{68\% joint confidence regions for $(\alpha,\beta)$ fitted 
using either $UB$ or $BV$ light curves of the test sample.
The crosses show the best fitted values with $1~\sigma$ uncertainties. 
\label{fig:contours}}
\end{figure}

We applied the same procedure as in the previous section to fit the 
$U$ and $B$-band light curves of the sub-sample of table~\ref{tab:test_sample} 
for which $U$-band measurements are available and redshifts larger than 0.015 (9 supernovae).
We obtain  $M_B^{70}= -19.37 \pm 0.05$, $\alpha= 0.8 \pm 0.4$, $\beta= 3.6 \pm 0.6$, 
and the standard deviation of residuals is $0.16 \pm 0.05$.
As expected,  
these results are consistent with the fit using $B$ and $V$, as 
shown by the confidence contours for $\alpha$ and $\beta$ fitted using either 
$UB$ or $BV$ light curves shown figure~\ref{fig:contours}. 
Note the covariance between the estimated values 
of $\alpha$ and $\beta$, particularly in the $U+B$ band case,
which simply reflects the correlation between the parameters $c$ and 
$s$ in the test sample.

One of the SN present in this sample (1996bo) show a large statistical uncertainty 
due to its limited $U$-band photometry. Removing this point from the fit has the 
effect of bringing down the $U,B$ contour 
to the point 
that it contains almost all of the $B,V$ contour, bringing the two determinations of      
$\alpha$ and $\beta$ closer to each other.  

\begin{figure}
\centering
\epsfig{figure=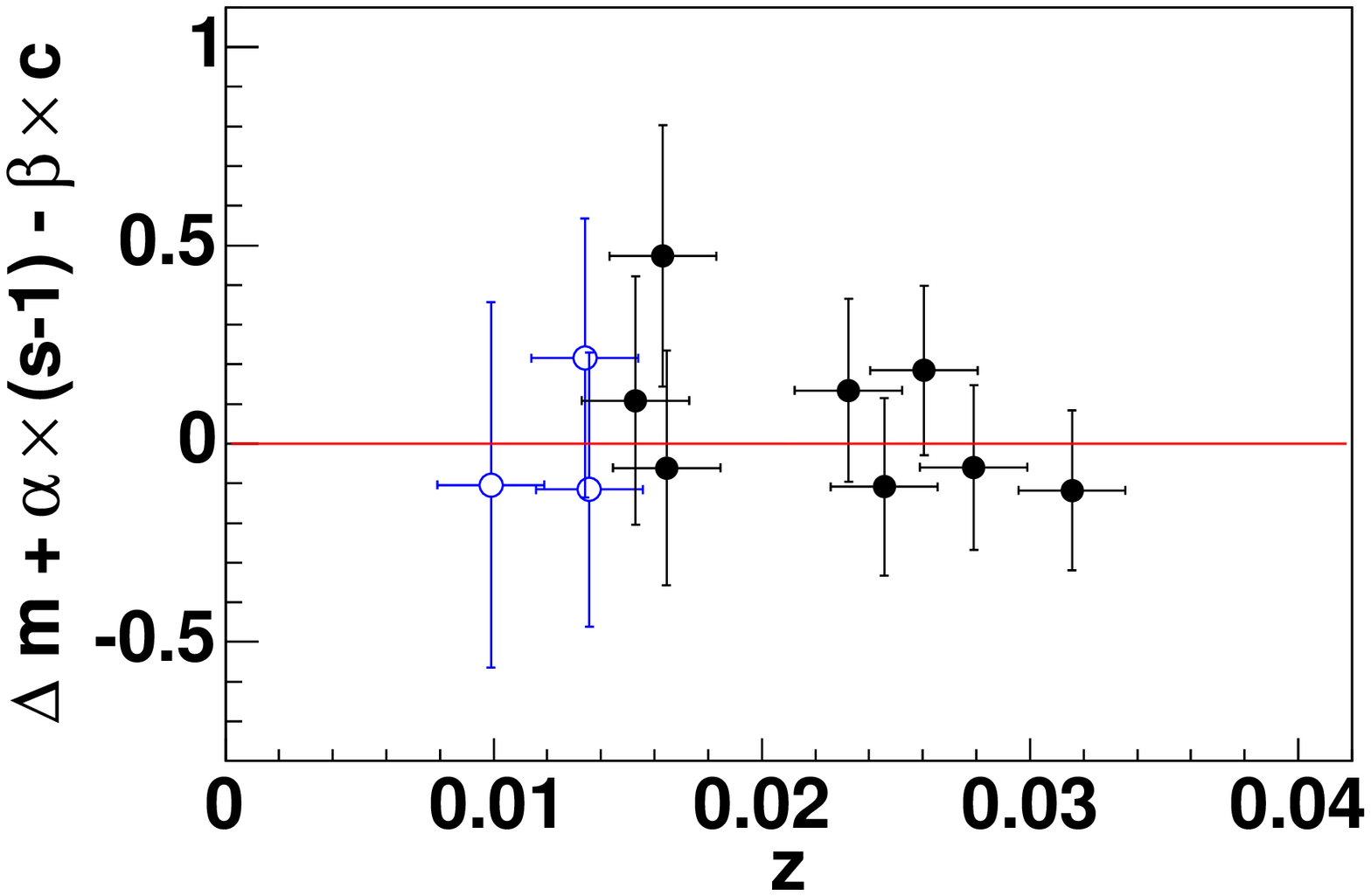,width=0.4\textwidth}
\epsfig{figure=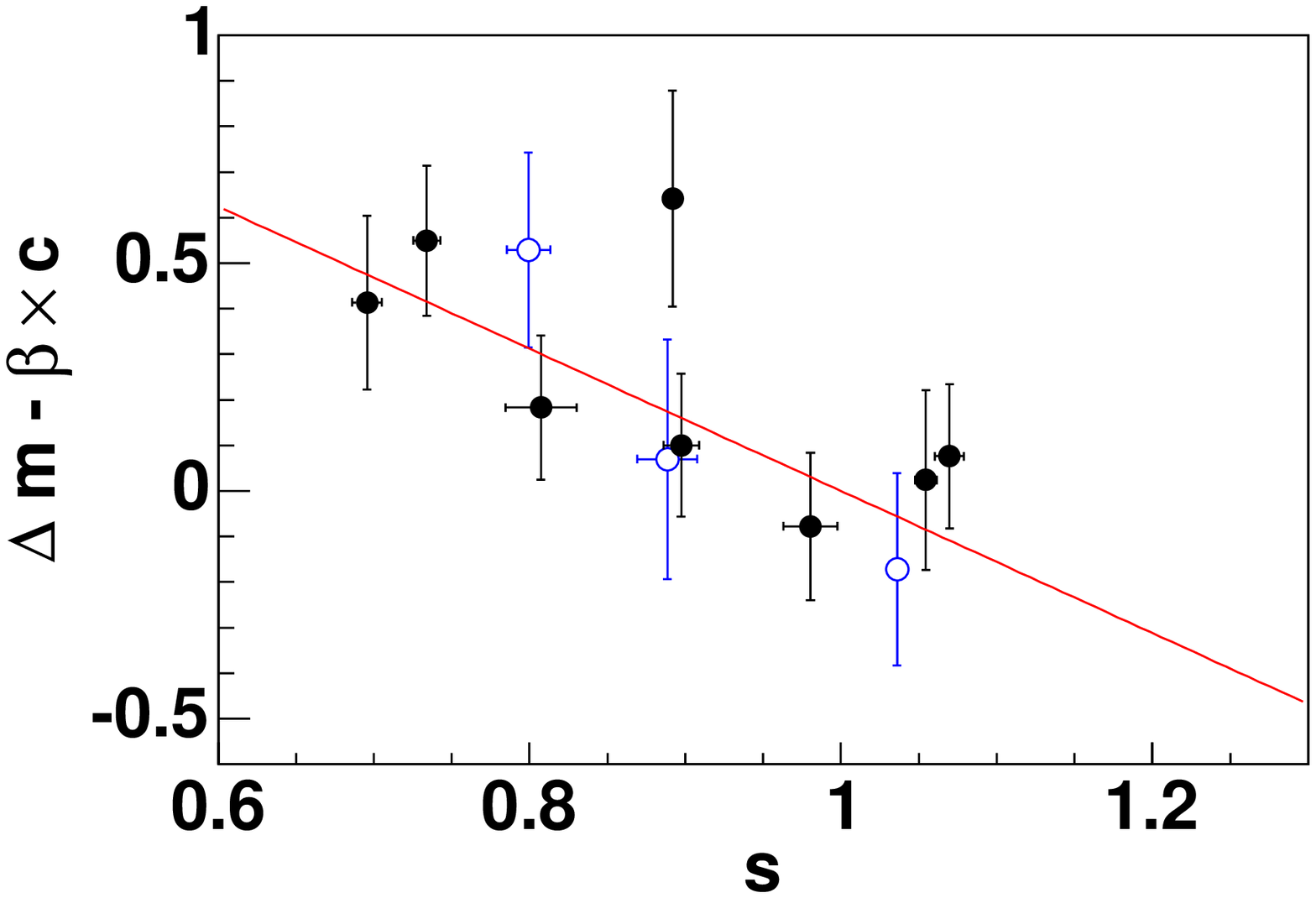,width=0.4\textwidth}
\epsfig{figure=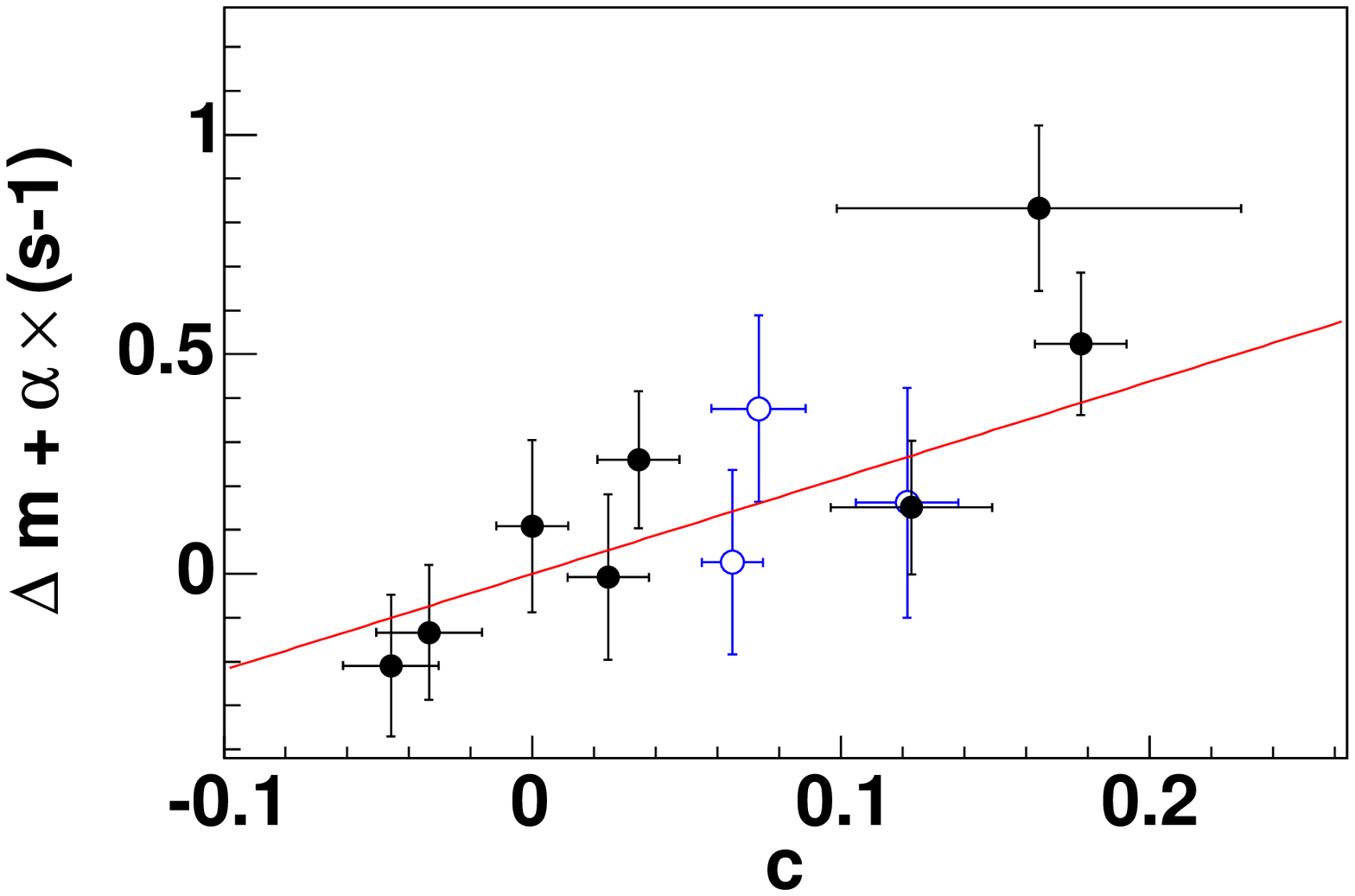,width=0.4\textwidth}
\caption{From top to bottom: Residuals to the Hubble diagram as a function 
of redshift, stretch $s$ and color $c$ indexes for supernovae of the test sample 
fitted in $U$ and $B$ bands. The values of $M'_B,\alpha,\beta$ used here are 
those fitted using $B$ and $V$ bands as described in the text. SNe with 
redshifts smaller than 0.015 are labeled with opened symbols.
Note the large uncertainty affecting one SN (1996bo) 
due to its limited $U$-band photometry, and which also 
appears as a $\sim2$ sigma outlier in the 3 plots. 
\label{fig:hubbleUB}}
\end{figure}

Figure~\ref{fig:hubbleUB} 
presents the residuals to the Hubble diagram as a function of redshift, stretch and 
color using the values of $M'_B,\alpha,\beta$ fitted with $B$- and $V$-band light 
curves in the previous section. Note the large uncertainty affecting 1996bo, which 
appears as a $\sim2$ sigma outlier in the 3 plots.
Fitting the Hubble diagram with the values of 
$\alpha$ and $\beta$ obtained with $B$- and $V$-band light curves, the standard 
deviation of residuals is $0.20 \pm 0.05$. 

\begin{figure}
\centering
\epsfig{figure=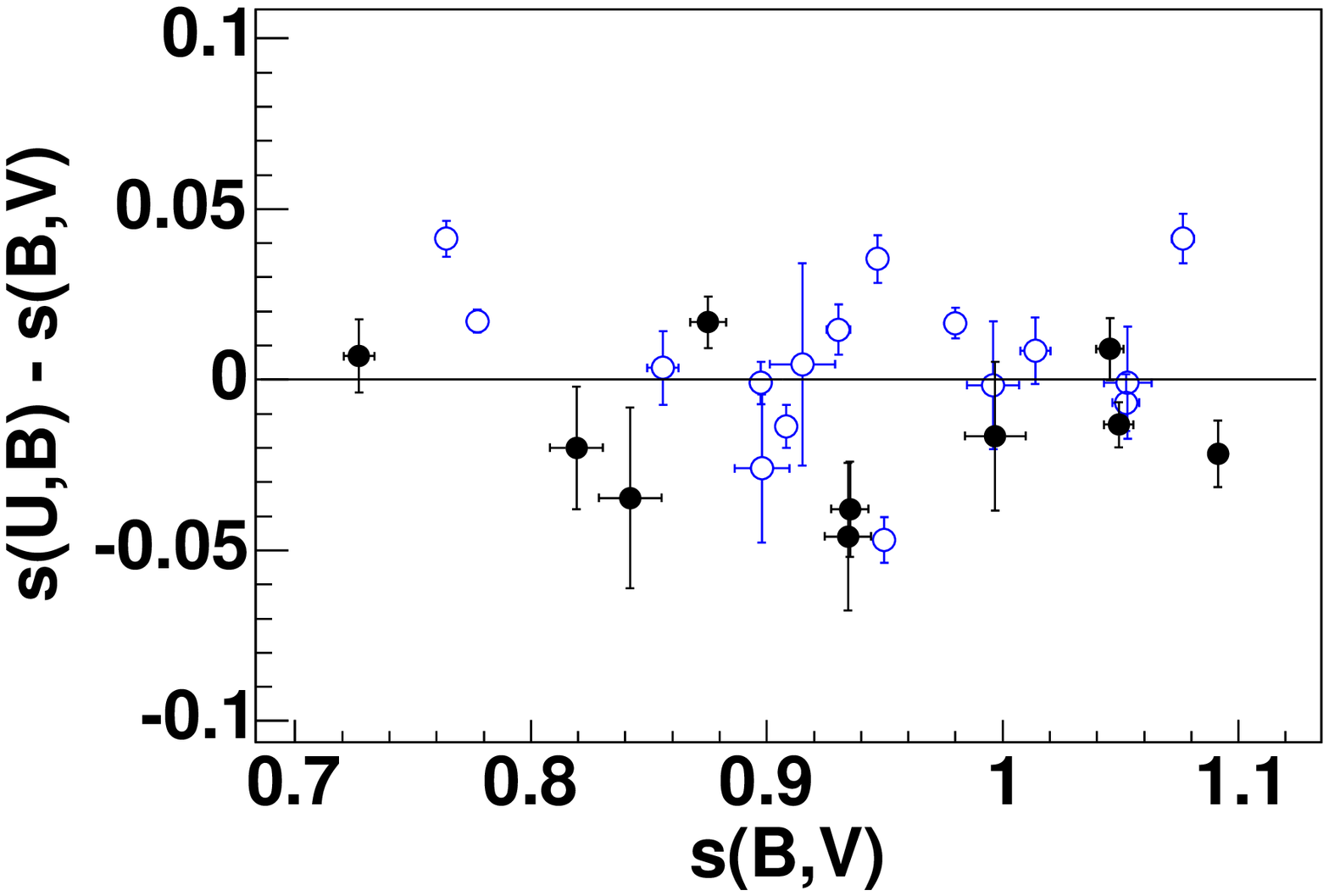,width=0.4\textwidth}
\epsfig{figure=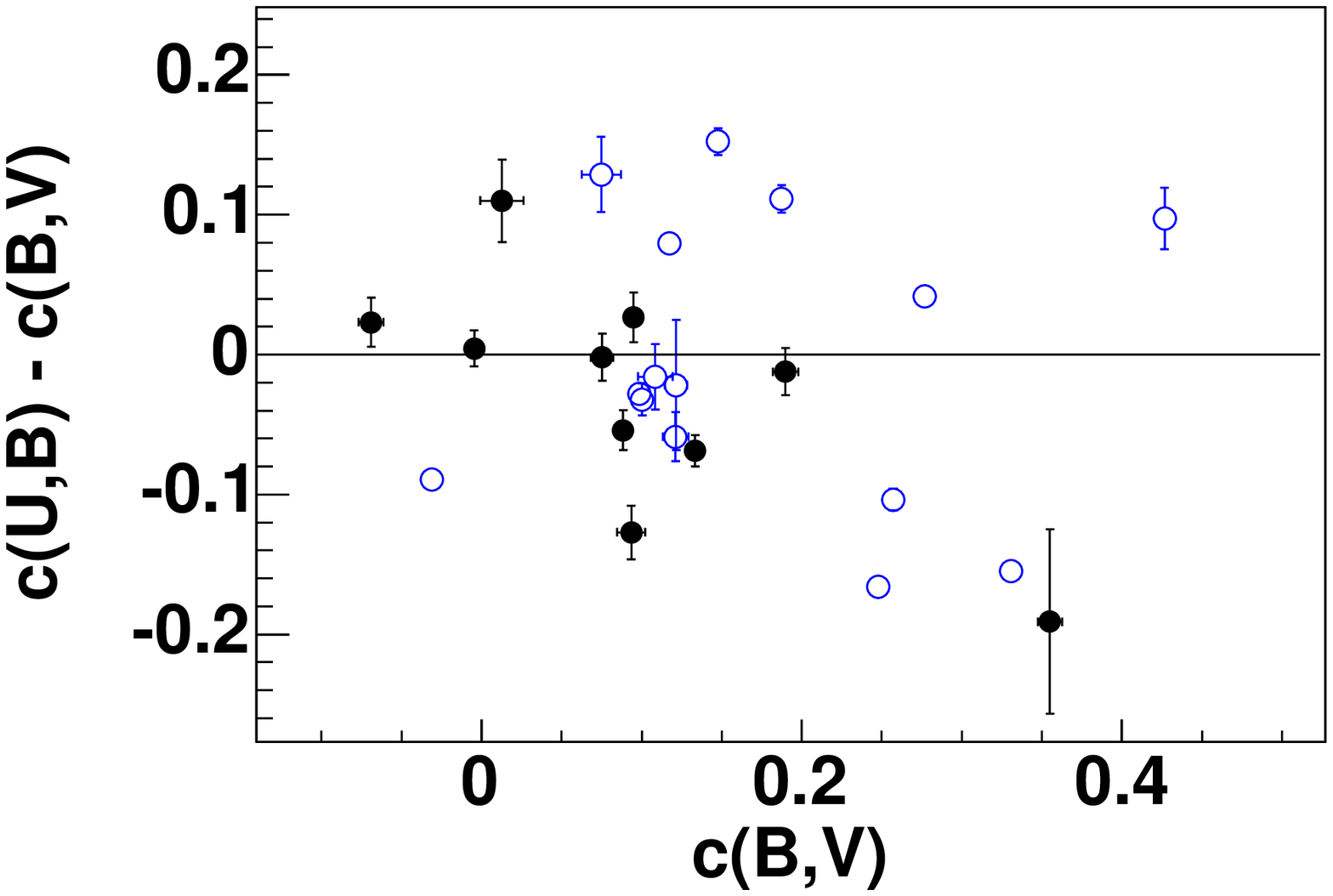,width=0.4\textwidth}
\caption{Differences of $s$ and $c$ values fitted with 
either $UB$ or $BV$ light curves of supernovae from the training and the test 
sample (respectively opened and filled symbols).
The error bars only reflect the propagation 
of photometric errors and do not account for any intrinsic dispersion.
\label{fig:comparisonUBV}}
\end{figure}

Comparisons of the fitted values of $s$ and $c$ fitted using either $UB$ 
or $BV$ data are shown figure~\ref{fig:comparisonUBV}. 
The error bars only reflect the propagation 
of photometric errors and do not account for any intrinsic dispersion.
There is no significant bias between the two estimates.   

\subsection{Comparison with other luminosity distance estimators}

We compare our method with various estimators of SN Ia distances
by examining the dispersions about the Hubble line of a sample of nearby 
supernovae. The comparison is made on distances measured in $B$ and $V$ bands only 
for which we obtain a dispersion of $0.16\pm0.03$. Our $U$- plus $B$-band distances 
estimates could not be compared with other estimate
due to the lack of published distances measured 
in these 2 bands.
  
As shown in many papers (e.g. \citealt{Wang03} and \citealt{Wang05}),
testing luminosity distance indicators on ``low-extinction sample''
greatly improves their performance. As said in the introduction, 
cutting on color estimate measured with a redshift
dependent accuracy is a source of systematic errors which we need
to avoid for
cosmological applications. We hence compare our resolutions to ``full-sample''
resolutions, for distance indicators involving rest-frame $U$, $B$ and $V$ bands.

The MLCS method was originally presented in \cite{Riess96a}, 
quoting a distance resolution of 0.12. Its  
latest development is presented
in \cite{Jha02}, with a distance resolution of 0.18, 
and used in \cite{Riess04}. 
From the latter, we collected the distance measurements
to the 20 objects in common with our test sample, and compute
a Hubble diagram dispersion of 0.24 (0.22 with the low-redshift
``golden sample'')
) to be compared with our value of 0.16 when measured on the 
same 20 events.

The CMAGIC method of \citet{Wang03} finds a weighted dispersion of 0.08
for a sub-sample of SNe with $B_{max}-V_{max}<0.05$.  With a weaker
cut on color, $B_{max}-V_{max}<0.5$, the dispersion rises to about
0.15, which is consistent with our result. Similarly, the 
$\Delta C_{12}$ method presented in \cite{Wang05}, also present an exquisite
distance resolution of 0.07 in the $V$ band based on a low-extinction subsample.
However, when considering the full sample, the distance resolution degrades
to 0.18.
A more detailed comparison with CMAGIC and $\Delta C_{12}$
would require comparing the distances of the same sample of SNe
(due to the limited statistics) but their distances 
are not published. 
Note also that in these papers,
the test and training samples are not separated.

In summary, the method we propose gives a dispersion on distances 
measured using $B$- and $V$-band data only comparable or lower than obtained
with other methods while also providing, for the first time, comparable dispersion 
values for distances measured using $U$- and $B$-band data only.

\section{Conclusion}

We have proposed a new method to fit broadband light curves of Type Ia
supernovae. It allows us to determine simultaneously the SN~Ia rest-frame $B$ magnitude
at maximum, stretch and color excess (or deficit) using any measured multi-color 
light curve within the wavelength range of rest-frame $UBV$ bands. 
This technique is particularly well-suited to the treatment of high-redshift 
SNe~Ia for which limited coverage is obtained in both wavelength and phase.  

The k-corrections, which allow the observer to
transform the observed magnitudes into the standard rest-frame magnitudes,  
are built-in; the model 
includes the dependence on stretch and color of the spectrum template needed
to estimate those corrections. In particular, the well-known correlation between 
$(U-B)_{max}$ and stretch is reproduced. 

The $(B-V)$ and $(V-R)$ stretch-independent colors
we obtain are extremely similar to the ones expected from reddening by dust. The $(U-B)$ 
color departs from this law. We find a relation between $(B-V)$ color and 
observed $B$ luminosity incompatible with $R_B = 4.1$, at more than 3 standard
deviations, even when accounting for the stretch--color correlation.

We have tested this fitting procedure on an independent sample 
of SNe~Ia. Alternatively using $B$- and $V$-band data and $U$- and $B$-band data, 
we are able to retrieve consistent parameters and hence build Hubble 
diagrams with both sets of data. The dispersions 
about the Hubble line were found to be $0.16 \pm 0.03$ and $0.16 \pm 0.05$ 
in the $B$ plus $V$ and $U$ plus $B$ bands only, respectively. 
This method is particularly well adapted  
to reliably measure SN Ia distances in the full redshift range 
and in particular beyond redshift $z\sim0.8$ for which rest-frame 
$V$-band measurements are often not available.

\begin{acknowledgements}
It is always a pleasure to acknowledge the stimulating discussions among the 
FROGS (FRench Observing Group of Supernovae), especially with G. Garavini, J. Rich 
and R. Taillet whom we would also like to thank for their critical reading 
of the manuscript. We thank P. Nugent for providing us with his template spectrum time series.
\end{acknowledgements}

\newpage
\bibliographystyle{aa}
\bibliography{bibi}

\clearpage
\onecolumn


\begin{deluxetable}{lccccccc}
\tablewidth{0pt}
\tablecaption{The training sample of Type Ia supernovae light curves}
\label{tab:training_sample}
\tablehead{
\colhead{Name} & 
\colhead{$z$ \tablenotemark{a}} & 
\colhead{Bands} &
\colhead{$m^*_B$} & 
\colhead{$s$} &
\colhead{$c$} &
\colhead{Phot. Ref.\tablenotemark{b}}  
}
\startdata
1981B & 0.006 & UBV & 12.018$\;$(0.003) & 0.911$\;$(0.004) & 0.165$\;$(0.002) & (B83) \cr
1984A & -0.001 & UBV & 12.389$\;$(0.003) & 0.946$\;$(0.005) & 0.215$\;$(0.003) & (K86) \cr
1986G & 0.002 & BV & 12.026$\;$(0.006) & 0.736$\;$(0.005) & 0.915$\;$(0.006) & (P87) \cr
1990N & 0.003 & UBVR & 12.701$\;$(0.006) & 1.055$\;$(0.005) & 0.090$\;$(0.005) & (L91,L98) \cr
1991T & 0.006 & UBVR & 11.574$\;$(0.003) & 1.129$\;$(0.005) & 0.183$\;$(0.003) & (F92,L98,A04,K04b) \cr
1992A & 0.006 & UBVR & 12.546$\;$(0.003) & 0.794$\;$(0.003) & 0.088$\;$(0.002) & (S92,A04) \cr
1992al & 0.015 & BVR & 14.448$\;$(0.012) & 0.922$\;$(0.011) & -0.035$\;$(0.012) & (H96) \cr
1994D & 0.001 & UBVR & 11.722$\;$(0.002) & 0.780$\;$(0.002) & -0.068$\;$(0.002) & (R95,P96,M96,A04) \cr
1994S & 0.015 & BVR & 14.764$\;$(0.017) & 1.006$\;$(0.025) & 0.011$\;$(0.018) & (R99) \cr
1994ae & 0.004 & BVR & 13.108$\;$(0.008) & 0.990$\;$(0.006) & 0.098$\;$(0.009) & (R99,A04) \cr
1995D & 0.007 & BVR & 13.450$\;$(0.013) & 1.029$\;$(0.014) & 0.072$\;$(0.012) & (R99,A04) \cr
1995al & 0.005 & BVR & 13.309$\;$(0.016) & 1.038$\;$(0.019) & 0.168$\;$(0.017) & (R99) \cr
1996X & 0.007 & BVR & 12.986$\;$(0.010) & 0.868$\;$(0.011) & 0.050$\;$(0.010) & (R99) \cr
1997E & 0.013 & UBVR & 15.101$\;$(0.006) & 0.820$\;$(0.010) & 0.078$\;$(0.006) & (J02) \cr
1997do & 0.010 & UBVR & 14.314$\;$(0.014) & 0.920$\;$(0.013) & 0.118$\;$(0.009) & (J02) \cr
1998bu & 0.003 & UBVR & 12.069$\;$(0.002) & 0.989$\;$(0.004) & 0.260$\;$(0.002) & (S99) \cr
1998dh & 0.009 & UBVR & 13.825$\;$(0.011) & 0.861$\;$(0.006) & 0.110$\;$(0.008) & (J02) \cr
1998es & 0.011 & UBVR & 13.814$\;$(0.007) & 1.061$\;$(0.009) & 0.104$\;$(0.006) & (J02) \cr
1999aa & 0.014 & UBVR & 14.728$\;$(0.005) & 1.055$\;$(0.005) & -0.007$\;$(0.005) & (J02,A04,K00) \cr
1999ac & 0.009 & UBVR & 14.114$\;$(0.005) & 0.925$\;$(0.009) & 0.112$\;$(0.006) & (J02) \cr
1999cc & 0.031 & UBVR & 16.802$\;$(0.009) & 0.834$\;$(0.012) & 0.047$\;$(0.010) & (J02) \cr
1999cl & 0.008 & UBVR & 14.819$\;$(0.008) & 0.901$\;$(0.011) & 1.118$\;$(0.007) & (J02,K00) \cr
1999dq & 0.014 & UBVR & 14.398$\;$(0.004) & 1.057$\;$(0.006) & 0.118$\;$(0.004) & (J02) \cr
1999ee & 0.011 & UBVR & 14.839$\;$(0.003) & 0.979$\;$(0.003) & 0.297$\;$(0.002) & (S02) \cr
1999ek & 0.018 & UBVR & 15.616$\;$(0.005) & 0.888$\;$(0.007) & 0.167$\;$(0.005) & (J02,K04b) \cr
2000E & 0.005 & UBVR & 12.856$\;$(0.004) & 1.011$\;$(0.006) & 0.219$\;$(0.004) & (V03) \cr
2000ca & 0.024 & UBVR & 15.529$\;$(0.007) & 1.000$\;$(0.012) & -0.066$\;$(0.006) & (K04a) \cr
2000cn & 0.023 & UBVR & 16.551$\;$(0.008) & 0.730$\;$(0.006) & 0.195$\;$(0.006) & (J02) \cr
2000dk & 0.017 & UBVR & 15.338$\;$(0.005) & 0.727$\;$(0.007) & 0.054$\;$(0.005) & (J02) \cr
2001V & 0.015 & BVR & 14.603$\;$(0.019) & 1.100$\;$(0.019) & 0.113$\;$(0.018) & (Vi03) \cr
2001bt & 0.014 & BVR & 15.267$\;$(0.006) & 0.865$\;$(0.005) & 0.232$\;$(0.007) & (K04b) \cr
2001cz & 0.016 & UBVR & 15.045$\;$(0.006) & 0.995$\;$(0.010) & 0.122$\;$(0.007) & (K04b) \cr
2001el & 0.004 & UBVR & 12.600$\;$(0.004) & 0.933$\;$(0.004) & 0.200$\;$(0.004) & (K03) \cr
2002bo & 0.004 & UBVR & 13.956$\;$(0.005) & 0.896$\;$(0.004) & 0.443$\;$(0.005) & (Z03,B04,K04b) \cr

\enddata
\tablenotetext{a}{Heliocentric redshift.}
\tablenotetext{b}{Photometry References : 
B83: \citet{Buta83}, 
K86: \citet{Kimeridze86}, 
P87: \citet{Phillips87}, 
L91: \citet{Leibundgut91},
F92: \citet{Filippenko92},
S92: \citet{Suntzeff92},  
R95: \citet{Richmond95}, 
H96: \citet{Hamuy96b}, 
M96: \citet{Meikle96}, 
P96: \citet{Patat96}, 
L98: \citet{Lira98}, 
R99: \citet{riess99a}, 
S99: \citet{Suntzeff99}, 
H01: \citet{Howell01}, 
L01: \citet{Li01b}, 
J02: \citet{Jha02}, 
S02: \citet{Stritzinger02}, 
K03: \citet{Krisciunas03},  
V03: \citet{Valentini03}, 
Vi03: \citet{Vinko03}, 
Z03: \citet{Zapata03}, 
A04: \citet{Altavilla04},  
B04: \citet{Benetti04}, 
G04: \citet{Garavini04}, 
K04b: \citet{Krisciunas04b}, 
G05: \citet{Garavini05}
}
\label{tab:training_sample}
\end{deluxetable}

\begin{deluxetable}{lcccccccccc}
\tablewidth{0pt}
\tablecaption{The test sample of Type Ia supernovae light curves}
\label{tab:test_sample}
\tablehead{
\colhead{Name} & 
\colhead{$z$ \tablenotemark{a}} & 
\colhead{$z$ \tablenotemark{b}} & 
\colhead{Bands} &
\colhead{$m^*_B~^{BV}$} & 
\colhead{$s~^{BV}$} &
\colhead{$c~^{BV}$} &
\colhead{$m^*_B~^{UB}$} & 
\colhead{$s~^{UB}$} &
\colhead{$c~^{UB}$} &
\colhead{Phot. Ref.\tablenotemark{c}}  
}
\startdata
1990af & 0.051 & 0.050  & BV & 17.73$\;$(0.01) & 0.73$\;$(0.01) & 0.00$\;$(0.01) & \nodata & \nodata & \nodata & (H96) \cr
1992bc & 0.020 & 0.020  & BVR & 15.09$\;$(0.01) & 1.03$\;$(0.01) & -0.04$\;$(0.01) & \nodata & \nodata & \nodata & (H96) \cr
1992bh & 0.045 & 0.045  & BV & 17.60$\;$(0.02) & 0.97$\;$(0.02) & 0.10$\;$(0.01) & \nodata & \nodata & \nodata & (H96) \cr
1992bo & 0.019 & 0.018  & BVR & 15.76$\;$(0.01) & 0.73$\;$(0.01) & 0.06$\;$(0.01) & \nodata & \nodata & \nodata & (H96) \cr
1992bp & 0.079 & 0.079  & BV & 18.28$\;$(0.01) & 0.86$\;$(0.01) & -0.04$\;$(0.01) & \nodata & \nodata & \nodata & (H96) \cr
1993H & 0.024 & 0.025  & BVR & 16.74$\;$(0.02) & 0.69$\;$(0.01) & 0.26$\;$(0.02) & \nodata & \nodata & \nodata & (H96,A04) \cr
1993O & 0.052 & 0.053  & BV & 17.62$\;$(0.01) & 0.89$\;$(0.01) & -0.01$\;$(0.01) & \nodata & \nodata & \nodata & (H96) \cr
1993ag & 0.049 & 0.050  & BV & 17.80$\;$(0.01) & 0.89$\;$(0.02) & 0.10$\;$(0.02) & \nodata & \nodata & \nodata & (H96) \cr
1995ac & 0.050 & 0.049  & BVR & 17.04$\;$(0.01) & 1.03$\;$(0.01) & 0.01$\;$(0.01) & \nodata & \nodata & \nodata & (R99,A04) \cr
1995bd & 0.016 & 0.016  & BVR & 15.26$\;$(0.01) & 0.99$\;$(0.01) & 0.30$\;$(0.01) & \nodata & \nodata & \nodata & (R99,A04) \cr
1996bl & 0.036 & 0.035  & BVR & 16.68$\;$(0.01) & 0.97$\;$(0.01) & 0.05$\;$(0.01) & \nodata & \nodata & \nodata & (R99) \cr
1996bo & 0.017 & 0.016  & UBVR & 15.83$\;$(0.01) & 0.88$\;$(0.01) & 0.36$\;$(0.01) & 15.84$\;$(0.01) & 0.89$\;$(0.01) & 0.16$\;$(0.07) & (R99,A04) \cr
1997E & 0.013 & 0.013  & UBVR & 15.10$\;$(0.01) & 0.82$\;$(0.01) & 0.08$\;$(0.01) & 15.10$\;$(0.01) & 0.80$\;$(0.01) & 0.07$\;$(0.02) & (J02) \cr
1998ab & 0.027 & 0.028  & UBVR & 16.08$\;$(0.01) & 0.94$\;$(0.01) & 0.09$\;$(0.01) & 16.05$\;$(0.02) & 0.90$\;$(0.01) & -0.03$\;$(0.02) & (J02) \cr
1999aa & 0.014 & 0.015  & UBVR & 14.73$\;$(0.01) & 1.05$\;$(0.01) & -0.00$\;$(0.01) & 14.72$\;$(0.01) & 1.05$\;$(0.01) & -0.00$\;$(0.01) & (J02,K00,A00) \cr
1999ac & 0.009 & 0.010  & UBVR & 14.10$\;$(0.01) & 0.93$\;$(0.01) & 0.09$\;$(0.01) & 14.08$\;$(0.01) & 0.89$\;$(0.02) & 0.12$\;$(0.02) & (J02) \cr
1999aw & 0.038 & 0.039  & BVR & 16.75$\;$(0.01) & 1.19$\;$(0.01) & 0.04$\;$(0.01) & \nodata & \nodata & \nodata & (S02) \cr
1999cc & 0.031 & 0.032  & UBVR & 16.78$\;$(0.01) & 0.84$\;$(0.01) & 0.01$\;$(0.01) & 16.75$\;$(0.02) & 0.81$\;$(0.02) & 0.12$\;$(0.03) & (J02) \cr
1999dq & 0.014 & 0.014  & UBVR & 14.41$\;$(0.01) & 1.05$\;$(0.01) & 0.13$\;$(0.01) & 14.41$\;$(0.01) & 1.04$\;$(0.01) & 0.06$\;$(0.01) & (J02) \cr
1999ek & 0.018 & 0.018  & BVR & 15.61$\;$(0.01) & 0.89$\;$(0.01) & 0.16$\;$(0.01) & \nodata & \nodata & \nodata & (J02,K04b) \cr
1999gp & 0.027 & 0.026  & UBVR & 16.02$\;$(0.01) & 1.09$\;$(0.01) & 0.09$\;$(0.01) & 16.02$\;$(0.01) & 1.07$\;$(0.01) & 0.03$\;$(0.01) & (J02,K01) \cr
2000ca & 0.024 & 0.025  & UBVR & 15.52$\;$(0.01) & 1.00$\;$(0.01) & -0.07$\;$(0.01) & 15.56$\;$(0.01) & 0.98$\;$(0.02) & -0.05$\;$(0.02) & (K04a) \cr
2000cn & 0.023 & 0.023  & UBVR & 16.55$\;$(0.01) & 0.73$\;$(0.01) & 0.19$\;$(0.01) & 16.56$\;$(0.01) & 0.73$\;$(0.01) & 0.18$\;$(0.01) & (J02) \cr
2000dk & 0.017 & 0.016  & UBVR & 15.34$\;$(0.01) & 0.73$\;$(0.01) & 0.05$\;$(0.01) & 15.33$\;$(0.01) & 0.70$\;$(0.01) & 0.02$\;$(0.01) & (J02) \cr
2001V & 0.015 & 0.016  & BVR & 14.60$\;$(0.02) & 1.10$\;$(0.02) & 0.11$\;$(0.02) & \nodata & \nodata & \nodata & (Vi03) \cr
2001ba & 0.029 & 0.031  & BV & 16.20$\;$(0.01) & 0.99$\;$(0.01) & -0.04$\;$(0.01) & \nodata & \nodata & \nodata & (K04a) \cr
\enddata
\tablenotetext{a}{Heliocentric redshift.}
\tablenotetext{b}{CMB-centric redshift.}
\tablenotetext{c}{Photometry References : 
H96: \citet{Hamuy96b},  
R99: \citet{riess99a},  
K00:  \citet{Krisciunas00}, 
H01: \citet{Howell01},  
L01: \citet{Li01b}, 
J02: \citet{Jha02},  
S02: \citet{Strolger02}, 
Vi03: \citet{Vinko03},  
Z03: \citet{Zapata03}, 
A04: \citet{Altavilla04}, 
G04: \citet{Garavini04}, 
K04a: \citet{Krisciunas04a},  
K04b: \citet{Krisciunas04b}, 
G05: \citet{Garavini05}
}
\label{tab:test_sample}
\end{deluxetable}

\end{document}